# Grid Added Value to Address Malaria

V. Breton[1], N. Jacq[1+2] and M. Hofmann[3]

[1]Laboratoire de Physique Corpusculaire, Université Blaise Pascal/IN2P3-CNRS UMR 6533, France
[2] Communication & Systèmes, CS-SI, France
[3] Fraunhofer Institute for Algorithms and Scientific Computing (SCAI), Department of Bioinformatics, Germany
*Breton,Jacq@clermont.in2p3.fr, martin.hofmann@scai.fraunhofer.de*

## Abstract

*Through this paper, we call for a distributed, internet-based collaboration to address one of the worst plagues of our present world, malaria. The spirit is a non-proprietary peer-production of information-embedding goods. And we propose to use the grid technology to enable such a world wide "open source" like collaboration. The first step towards this vision has been achieved during the summer on the EGEE grid infrastructure where 46 million ligands were docked for a total amount of 80 CPU years in 6 weeks in the quest for new drugs.*

## 1. Introduction

The number of cases and deaths from malaria increases in many parts of the world. There are about 300 to 500 million new infections, 1 to 3 million new deaths and a 1 to 4% loss of gross domestic product (at least $12 billion) annually in Africa caused by malaria.

The main causes for the comeback of malaria are that the most widely used drug against malaria, chloroquine, has been rendered useless by drug resistance in much of the world [17] and that anopheles mosquitoes, the disease vector, have become resistant to some of the insecticides used to control the mosquito population.

Genomics research has opened new ways to find novel drugs to cure malaria, vaccines to prevent malaria, insecticides to kill infectious mosquitoes and strategies to prevent development of infectious sporozoites in the mosquito [2]. These studies require more and more in silico biology; from the first steps of gene annotation via target identification to the modelling of pathways and the identification of proteins mediating the pathogenic potential of the parasite. Grid computing supports all of these steps and, moreover, can also contribute significantly to the monitoring of ground studies to control malaria and to the clinical tests in plagued areas.

It is well understood that the grid itself cannot drive drug development, but it can function as the catalyst that brings the actors (biochemists, physicians, etc) together and pushes them ahead in the same direction.

## 2. Genomics research areas on malaria

To increase the chances to develop new and better drugs and vaccines, it is very important to know the sequence of the genomes of the parasites that cause malaria as well as the specifics of gene and protein expression at different stages in the life cycle and under pressure from different drugs.

The availability of genomic information for Anopheles gambiae, the major vector in Africa, and Plasmodium spec. enables us now to work out new approaches to interfere with the mechanism of the development of infectious sporozoites in anopheles mosquitoes and for reducing contact between infectious mosquitoes and humans. Much effort is also being devoted to biochemical and molecular studies of resistance mechanisms and to mapping of resistance genes.

### 2.1. Development of vaccines

In the development of vaccines, the value of knowing the sequences of certain genes and/or peptides is undeniable. It is known that susceptibility to Plasmodium can be entirely prevented for 9 months by immunizing with radiation attenuated sporozoites [1] and those children who live to 3 to 10 years of age in areas endemic for malaria rarely if ever develop severe malaria and die. If protective immunity requires induction of immune responses against 100 or even

500 different target proteins, genomics research offers the most immediate way to identify those targets and to begin the process of developing vaccines based on them.

However, progress has been painfully slow, and the few months duration of protection from the best currently available vaccine is inferior to what can be achieved with insecticide-treated nets or house spraying. When better vaccines eventually emerge, they should probably be used in conjunction with vector control to avoid their effects being hidden in areas of intense transmission. Whether vaccines and vector control synergize will have to be tested in the field and will not be predictable from molecular properties.

## 2.2. Human susceptibility to malaria

Knowledge of the human genome offers unprecedented potential for understanding who is and is not susceptible to dying from malaria, and who can benefit most from a particular type of vaccine. Having the human and parasite blueprint, and the computational biology capacity that goes with it, may allow developing a simple diagnostic that would identify at birth who is most at risk of dying from malaria. A single nucleotide polymorphism in the gene encoding the beta chain of human haemoglobin is associated with a 90% decrease in the chance of dying from a Plasmodium falciparum infection. Genomics research allows exploring if other SNPs or SNPs complexes in the human, anopheles and parasite genomes could impact the infection and the disease.

## 2.3. Transgenesis

There is also interest about transgenesis as a way to generate strains of mosquito that cannot transmit malaria. However, without extremely reliable systems for driving the transgenes into wild sector populations, possession of a non-transmitter strain would be of no practical use.

A more feasible and acceptable way of using transgenesis may be to improve the sterile insect technique, which would only require release of non-biting males.

## 3. Rationale for a grid to address malaria

For both vector control and chemotherapy, knowing the gene sequences of Anopheles and Plasmodium species should lead to discovery of targets against which new insecticides or antimalarial drugs can be produced. Genomics research is crucial to attacking malaria, but this research must be strongly coupled to ground studies in order to guide disease controllers, especially those working in countries with annual health budgets of less than $10 per person. Evaluation of the impact of new drugs and new vaccines require careful monitoring of clinical tests, especially in areas of high malaria transmission where it is important to distinguish recurrence of parasites, due to recrudescence of incompletely cured infections, from reinfection due to new mosquito bites. Disease controllers are also faced with the necessity to monitor goal-oriented field work on a long term.

The grid technology provides the collaborative IT environment to enable the coupling between molecular biology research and goal-oriented field work. It proposes a new paradigm for the collection and analysis of distributed information where data are no more to be centralized in one single repository. On a grid, data can be stored anywhere and still be transparently accessed by any authorized user. The computing resources of a grid are also shared and can be mobilised on demand so as to enable very large scale genomics comparative analysis and virtual screening.

The motivating perspective is to enhance the ability of both pharmaceutical industry and academic research institutions to share diverse, complex and distributed information on a given disease for collaborative exploration and mutual benefit. Leading international pharmaceutical groups such as Pfizer, Sanofi-Aventis and Novartis are already involved in activities of the United Nations / WHO to combat diseases of the poor. The goal here is to lower the barrier to such substantive interactions in order to produce cheaper drugs and insecticides to address diseases affecting third world development and to increase the return on investment for new drugs in the developed countries.

## 4. Grid added value to address malaria

### 4.1. *In silico* molecular biology

Genomics research areas on malaria were discussed previously and include:

- Search for targets on *Anopheles* genome and proteome for new insecticides
- Search for targets on *Plasmodium* genomes and proteomes for new

- drugs (e.g. inhibitors of parasite development)
- Identification of target proteins on *Plasmodium* proteomes to induce sustainable immune responses
- Study of human susceptibility to malaria : identification of SNPs on human and *Plasmodium* genomes related to sensitivity
- Study of *Plasmodium* and *Anopheles* genomics to prevent development of infectious sporozoites in anopheles mosquitoes
- Study of mechanisms for drug resistance; including distributed monitoring of epidemiological parameters
- Transgenesis to generate strains of mosquitoes that cannot transmit malaria

These studies require extensive computing and storage resources.

### 4.2. *In silico* drug discovery

Virtual screening is about selecting in silico the best candidate drugs acting on a given target protein. Screening can be done in vitro using real chemical compounds, but this is a very expensive undertaking. If it could be done in silico in a reliable way, one could reduce the number of molecules requiring in vitro and then in vivo testing from a few millions to a few hundreds [15]. Advance in combinatorial chemistry has paved the way for synthesizing millions of different chemical compounds. Thus there are millions of chemical compounds available in pharmaceutical laboratories and also in a very limited number of publicly accessible databases.

A coordinated public effort on in silico drug screening would require to set up a "public" collection of virtual compounds from existing compound databases and compound catalogues (PDB [9] Ligand Chemistry, KEGG-Ligand [10], compound catalogues, PubChem [21]…). This public compound repository should also comprise all information known about biological activities of the compounds in this repository. Moreover, information on possible strategies for the synthesis of these compounds should be available to speed up the in vitro testing of potential candidate drugs after in silico screening.

In silico drug discovery should foster collaboration between public and private laboratories. It should also have an important societal impact by lowering the barrier to develop new drugs for rare and neglected diseases.

### 4.3. Federation of databases for clinical tests in plagued areas

Grid technology opens new perspectives for preparation and follow-up of medical missions in developing countries as well as support to local medical centres in terms of teleconsulting, telediagnosis, patient follow-up and e-learning [13]. In every hospital, patients are recorded in the databases which are federated in a network. Such a federation of databases allows medical data to be kept distributed in the hospitals behind firewalls. Views of the data are granted according to individual access rights through secured networks.

Such a federation of databases can be used to monitor ground studies in developing countries. The data stored can be used for epidemiology purposes as well as clinical tests.

In testing antimalarial drugs in areas of high malaria transmission, it is important to distinguish recurrence of parasites, due to recrudescence of incompletely cure infections, from reinfection due to new mosquito bites. Molecular matching of the parasite clones in the same individuals before and after treatment is a way of doing this.

When better vaccines eventually emerge, they should probably be used in conjunction with vector control to avoid their effects being hidden in areas of intense transmission. Whether vaccines and vector control synergize will have to be tested in the field and will not be predictable from molecular properties.

### 4.4. Knowledge space for disease information

The goal is to make all relevant information on the disease available to all interested parties. The concept of a knowledge space is to organize the information so that it can be reached in a few clicks. This concept is already successfully used internally by pharmaceutical laboratories to store knowledge [14]. The grid allows building a distributed knowledge space so that each participant is able to keep the information he owns on his local computer. A set of grid services is proposed to make information easily consultable for the different potential customers (physicians, researchers, etc.). These services would take advantage of the developments in the area of semantic text analysis and text mining for extraction of information in biology and genome research. Moreover, a structured knowledge space would produce grid services for indexing of distributed data resources and thus improve navigation through knowledge and retrieval of relevant information.

### 4.5. Monitoring of fieldwork to control malaria

Monitoring tools to control malaria in plagued areas include reducing contacts between infected mosquitoes and humans by filling in breeding sites, larviciding, spraying houses with insecticides, insecticide-impregnated bed nets, and house screening. Effective application of these measures would lead to reduce morbidity and mortality from malaria. However, given the extremely high transmission rate of Plasmodium falciparum, especially where Anopheles gambiae mosquitoes predominate, the impact of these tools can be limited by the capacity of mosquitoes to develop resistance and by the requirement for maintenance of the interventions for many years.

Long term monitoring of these tools clearly would benefit of technologies allowing a better collection and analysis of distributed data in developing countries as is the case with grids.

Indeed, internet is now accessible worldwide so the idea is to organize the collect of information around regional centres acting as local repositories. These repositories would be federated thanks to the grid technology at a world level for a general overview of the vector control work at the level of international agencies or non-profit organizations.

## 5. Technology requirements for a malaria grid

The goal of a grid for malaria would be to handle all aspects of the fight against malaria:

- Search of new drug targets through post-genomics requiring data management and computing
- Massive docking to search for new drugs requiring high performance computing and data storage
- Handling of clinical tests and patent data requiring data storage and management
- Overseeing the distribution of the existing drugs requiring data storage and management
- Overseeing of vector control requiring data storage and management

The grid should gather:

- Biomedical laboratories searching for vaccines, working on the genomes of the parasite and/or the parasite vector and/or the human genome
- Drug designers to identify new drugs
- Healthcare centres involved in clinical tests
- Healthcare centres collecting patent information
- Structures involved in distributing existing treatments (healthcare administrations, non profit organizations…) and enforcing vector control
- IT technology developers and computing centres

The grid should provide the following services:

- Large computing and storage resources for genomics research and virtual docking
- Access to relevant genomes, regularly updated data bases and publications
- Knowledge space with genomics and medical information (epidemiology, status of clinical tests, drug resistances, etc.)
- Collaboration environment for the participating partners. No one entity can have an impact on all R&D aspects involved in addressing one disease as complex as malaria. The grid would act as a virtual laboratory of the different actors.
- Federation of regional data bases for monitoring of vector control and clinical tests

## 6. An open source approach to drug discovery

The proposed grid to address malaria borrows the "open source" approach that has proven so successful in software development. Open source approaches have emerged in the biotechnology already. The international effort to sequence the human genome for instance resembled an open source initiative. It placed all the resulting data into the public domain rather than allow any participant to patent any of the results.

The open source research on malaria could be organized as follows: a web site would allow chemists and biologists to volunteer their expertise on certain areas of the disease. They would examine and annotate shared databases and perform experiments. The results would be fully transparent and discussed in chat rooms. The research would be initially mainly computational, based on resources provided by grid infrastructures, and not carried out in "wet" laboratories.

The difference between this proposal and earlier open source approaches in biomedical research is that scientists would collaborate on the data and not only on the software.

The final development of drug candidates could be awarded to a laboratory based on competitive bids. The drug itself would go to public domain, for generic manufacturers to produce. This would achieve the goal of getting new medicines to those who need them at the lowest possible price. This model is currently supported by various programs of the United Nations and involves several commercial organisations, including large pharma companies.

Open source research could also open the area of non-patentable compounds and drugs whose patents have expired. These receive very little attention from researchers because there would be no way to protect and so profit from any discovery that was made about their effectiveness. Lots of potentially useful drugs could be sitting under researchers noses.

## 7. WISDOM, first step towards grid-enabled virtual screening

*As* discussed previously, in silico drug discovery is one of the most promising strategies to speed-up the drug development process. High throughput virtual screening allows screening millions of compounds rapidly, reliably and in a cost effective way.

Grids like EGEE [4] are ideally suited for the first docking step where docking probabilities are computed for millions of ligands. It has been demonstrated during the summer 2005 by the WISDOM [16] initiative on malaria where 46 million ligands were docked for a total amount of 80 CPU years in 6 weeks.

### 7.1. EGEE - an enabling grid for eScience

The EGEE project [4] (Enabling Grid for E-sciencE) brings together experts from over 27 countries with the common aim of building on recent advances in Grid technology and developing a service Grid infrastructure which is available to scientists 24 hours-a-day. The project aims to provide researchers in academia and industry with access to major computing resources, independent of their geographic location. The EGEE infrastructure is now a production grid with a large number of applications installed and used on the available resources. The infrastructure involves more than 180 sites spread in Europe, America and Asia.

The WISDOM application was deployed within the framework of the biomedical Virtual Organization (VO). Figure 1 gives an overview of the geographical distribution of the resources node available for biomedical applications, which scale up to 3000 CPUs and 21 TB disk space.

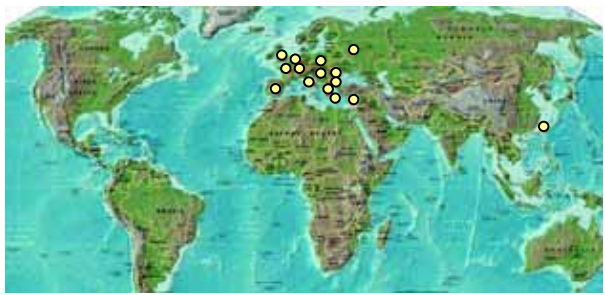

**Figure 1: Geographical distribution of EGEE biomedical resources in August 2005. The resources are spread over 15 countries.**

### 7.2. The WISDOM initiative

WISDOM stands for World-wide In Silico Docking On Malaria. Its goal was to propose new inhibitors for a family of proteins produced by Plasmodium falciparum using in silico virtual docking on a grid infrastructure. The chosen target is involved in the haemoglobin metabolism, which is one the key metabolic processes in the survival of the parasite.

There are several proteases involved in human haemoglobin degradation inside the food vacuole of the parasite inside the erythrocytes. Plasmepsin, the aspartic protease of Plasmodium, is responsible for the initial cleavage of human haemoglobin and later followed by other proteases [18]. There are ten different plasmepsins coded by ten different genes in Plasmodium falciparum (Plm I, II, IV, V, VI, VII,

VIII, IX, X and HAP) [19]. High levels of sequence homology are observed between different plasmepsins (65-70%). Simultaneously they share only 35% sequence homology with its nearest human aspartic protease, Cathepsin D4 [20]. This and the presence of accurate X crystallographic data make plasmepsins ideal targets for rational drug design against malaria.

Two docking software packages were used in WISDOM: one is FlexX [6], a commercial software made graciously available by BioSolveIT [12] for a limited time, and the other is Autodock [7], a software which is open-source for academic laboratories and which uses a different docking method. Chemical compounds were obtained from the ZINC database [8]. A subset of the ZINC database, the ChemBridge database (~500000 compounds), was docked using FlexX and Autodock. Another drug like subset (~500000 compounds) of ZINC database was docked using FlexX.

### 7.3. WISDOM - results and perspectives

During 6 weeks, 72751 jobs were launched for a total of 80 CPU years, producing 1 TB of data. A record number of 1643 jobs ran in parallel on 58 grid nodes, each representing from a few to more than 1000 processors [5].

First analysis of WISDOM results at Fraunhofer Institute (SCAI) has selected the 10,000 most promising compounds out of 1 million candidates using docking scores. The best scoring ligands include both known inhibitors and new ones. Most important is, that besides well established chemical core structures that share features with the already known plasmepsin inhibitors, new chemical entities were discovered. This is a first demonstration of the relevance of the in silico approach.

Before in vitro experimentation, the best compounds will be re-ranked using molecular modelling within the framework of the BioInfoGrid European grid project. In parallel, a new massive deployment of in silico docking is foreseen on the EGEE infrastructure with new targets in the fall of 2006.

## 8. Conclusion

Through this paper, we call for a distributed, internet-based collaboration to address one of the worst plagues of our present world, malaria. The spirit is a non-proprietary peer-production of information-embedding goods. And we propose to use the grid technology to enable such a world wide "open source" like collaboration. The first step towards this vision has been achieved during the summer on the EGEE grid infrastructure where 46 million ligands were docked for a total amount of 80 CPU years in 6 weeks in the quest for new drugs. Further development of a grid-enabled virtual screening pipeline is underway in several European projects.

## 9. Acknowledgements

Many ideas expressed in this document were inspired by reading [2] and [3] as well as discussions with H. Bilofsky, C. Jones, M. Peitsch, T. Schwede and R. Ziegler. EGEE is a project funded by the European Union under contract INFSO-RI-508833.

## 10. References

[1] R. Elderman et al, J. Infect. Dis. 168, 1066. (1993)
[2] C.F Curtis and S.L. Hoffman, Science 290, 1508-1509 (2000)
[3] An open-source shot in the arm ? The Economist Technology Quarterly, June 12th 2004
[4] Fabrizio Gagliardi, Bob Jones, François Grey, Marc-Elian Bégin, Matti Heikkurinen, "Building an infrastructure for scientific Grid computing: status and goals of the EGEE project". Philosophical Transactions: Mathematical, Physical and Engineering Sciences, Issue: Volume 363, Number 1833 / August 15, 2005, Pages: 1729 – 1742, DOI:10.1098/rsta.2005.1603
[5] N. Jacq, J. Salzemann, Y. Legré, M. Reichstadt, F. Jacq, E. Medernach, M. Zimmermann, A. Maas, M. Sridhar, K. Vinodkusam, J. Montagnat, H. Schwichtenberg, M. Hofmann and V. Breton, In silico docking on grid infrastructures: the case of WISDOM, submitted to FGCS, 2006
[6] M. Rarey, B. Kramer, T. Lengauer, G. Klebe, Predicting Receptor-Ligand interactions by an incremental construction algorithm, J. Mol. Biol. 261 (1996) 470-489.
[7] G.M. Morris, D.S. Goodsell, R.S. Halliday, R. Huey, W.E. Hart, R.K. Belew, A.J. Olson, Automated Docking Using a Lamarckian Genetic Algorithm and Empirical Binding Free Energy Function, J. Computational Chemistry, 19 (1998) 1639-1662.
[8] Irwin, Shoichet, J. Chem. Inf. Model. 45(1) (2005) 177-82.
[9] H.M. Berman, J. Westbrook, Z. Feng, G. Gilliland, T.N. Bhat, H. Weissig, I.N. Shindyalov, P.E. Bourne, The Protein Data Bank, Nucleic Acids Research, 28 (2000) 235-242.
[10] Kanehisa, M., Goto, S., Hattori, M., Aoki-Kinoshita, K.F., Itoh, M., Kawashima, S., Katayama, T., Araki, M., and Hirakawa, M.; From genomics to chemical genomics: new developments in KEGG. Nucleic Acids Res. 34, D354-357 (2006)
[11] Kuntz, I. D.; Blaney, J. M.; Oatley, S. J.; Langridge, R.; Ferrin, T. E. A Geometric Approach to Macromolecule-Ligand Interactions. J. Mol. Biol. 1982, 161, 269-288
[12] BioSolveIT Homepage: http://www.biosolveit.de

[13] Florence Jacq, Frank Bacin, Nonfounikoun Meda, Denise Donnarieix, Jean Salzemann, Vincent Vayssiere, Michel Renaud, François Traore, Gertrude Meda, Rigobert Nikiema and Vincent Breton, Towards grid-enabled telemedicine in Africa, submitted to IST Africa 2006 conference

[14] M. C. Peitsch et al., Informatics and knowledge management at the Novartis Institutes for BioMedical Research, SCIP-online, 46 (2004) 1-4.

[15] R.W. Spencer, Highthroughput virtual screening of historic collections on the file size, biological targets, and file diversity, Biotechnol. Bioeng 61 (1998) 61-67.

[16] WISDOM Homepage : http://wisdom.eu-egee.fr

[17] J. Weisner, R. Ortmann, H. Jomaa, M. Schlitzer, Angew. New Antimalarial drugs, Chem. Int. 42 (2003) 5274-529.

[18] S. E. Francis, D. J. Jr. Sullivan, D.E. Goldberg, Hemoglobin metabolism in the malaria parasite plasmodium falciparum, Annu.Rev. Microbiol. 51 (1997), 97-123.

[19] G. H. Coombs, D.E. Goldberg, M. Klemba, C. Berry, J. Kay, J.C. Mottram, Aspartic proteases of plasmodium falciparum and other protozoa as drug targets, Trends parasitol. 17 (2001), 532-537.

[20] A.M. Silva, A.Y. Lee, S.V. Gulnik, P. Majer, J. Collins, T.N. Bhat, P.J. Collins, R.E. Cachau, K.E. Luker, I.Y. Gluzman, S.E. Francis, A. Oksman, D.E. Goldberg, J.W. Erickson, Structure and inhibition of plasmepsin II, A haemoglobin degrading enzyme from Plasmodium falciparum, Proc. Natl. Acad. Sci. USA 93 (1996) 10034-10039.

[21] PubChem, http://pubchem.ncbi.nlm.nih.gov/